\documentclass[twocolumn,9pt]{extarticle}

\usepackage{amsmath, amsfonts}
\RequirePackage{lmodern}
\usepackage[T1]{fontenc}
\usepackage{graphicx}
\usepackage{dcolumn}
\usepackage{bm}
\usepackage[mathlines]{lineno}
\usepackage[lmargin=48.5pt, rmargin=53.5pt, tmargin=40pt, bmargin=60pt, columnsep=11.5pt]{geometry}
\usepackage{caption}
\usepackage{titling}
\usepackage[superscript, biblabel]{cite}
\usepackage{physics}
\usepackage{setspace}

\usepackage{etoolbox}
\patchcmd\thebibliography
 {\labelsep}
 {\labelsep\itemsep=-1.9pt}
 {}
 {\typeout{Couldn't patch the command}}
 
\makeatletter
\renewcommand*{\@fnsymbol}[1]{\ensuremath{
	\ifcase#1\or \textsf{1} \or \dag \or \star \or \textsf{2} \else\@ctrerr\fi}}
\renewcommand\@cite[2]{%
Ref.~#1\ifthenelse{\boolean{@tempswa}}
{, \nolinebreak[3] #2}{}
}
\renewcommand\@biblabel[1]{#1.}
\makeatother

\newcommand*\samethanks[1][\value{footnote}]{\footnotemark[#1]}

\pretitle{\begin{flushleft}}
\posttitle{\end{flushleft}}
\preauthor{\begin{flushleft}}
\postauthor{\end{flushleft}}
 
\begin{document}

\title{\textsf{\textbf{{\fontsize{25}{60}\selectfont Signatures of two-photon pulses from a quantum two-level system}}}}

\author{\large\textbf{\textsf{Kevin A. Fischer\thanks{E. L. Ginzton Laboratory, Stanford University, Stanford CA 94305, USA}\hspace{4pt}\thanks{These authors contributed equally}\hspace{4pt}\thanks{email: kevinf@stanford.edu; kai.mueller@wsi.tum.de}\hspace{3pt}, Lukas Hanschke\thanks{Walter Schottky Institut and Physik Department, Technische Universit\"at M\"unchen, 85748 Garching, Germany}\hspace{4pt}\samethanks[2]\hspace{3pt}, Jakob Wierzbowski\samethanks[4]\hspace{3pt}, Tobias Simmet\samethanks[4]\hspace{3pt}, Constantin Dory\samethanks[1]\hspace{3pt}, ~~~~~~~~~~~~ Jonathan J. Finley\samethanks[4]\hspace{3pt}, Jelena Vu\v{c}kovi\'c\samethanks[1],  Kai M\"uller\samethanks[4]\hspace{4pt}\samethanks[3]\vspace{-1.5ex}}}}

\date{\vspace{-4ex}}

\begingroup
\let\center\flushleft
\let\endcenter\endflushleft
\maketitle
\endgroup

\begin{spacing}{0.97}

\noindent\textsf{\textbf{The theoretical community has found interest in the ability of a two-level atom to generate a strong many-body interaction with light under pulsed excitation \cite{shen2007sc,zheng2010waveguide,pletyukhov2015quantum}. Single-photon generation is the most well-known effect, where a short Gaussian laser pulse is converted into a Lorentzian single-photon wavepacket \cite{schneider2015single,Fischer2016-pl}. However, recent proposals have surprisingly suggested that scattering with intense laser fields off a two-level atom may generate oscillations in two-photon emission that are out of phase with its Rabi oscillations, as the power of the pulse increases \cite{pletyukhov2013full,lindkvist2014scattering}. Here, we provide an intuitive explanation for these oscillations using a quantum trajectory approach \cite{carmichael2009open} and show how they may preferentially result in emission of two-photon pulses. Experimentally, we observe signatures of these oscillations by measuring the bunching of photon pulses scattered off a two-level quantum system. Our theory and measurements provide crucial insight into the re-excitation process that plagues \cite{dada2016indistinguishable,Fischer2016-pl} on-demand single-photon sources while suggesting the production of novel multi-photon states.
}
}

\textit{Theory}---We begin by considering an ideal quantum two-level system that interacts with the outside world only through its electric dipole moment $\mu$ \cite{Cohen-Tannoudji1992-uo}. Suppose the system is instantaneously prepared in the superposition of its ground $\ket{g}$ and excited $\ket{e}$ states
\begin{equation}
\ket{\psi_i}=\sqrt{1-\textrm{P}_e}\ket{g}+\sqrt{\textrm{P}_e}\ket{e},
\end{equation}
where $\textrm{P}_e$ is the probability of initializing the system in $\ket{e}$. From this point, spontaneous emission at a rate of $\Gamma$ governs the remaining system dynamics and a single photon is coherently emitted with probability $\textrm{P}_e$, while no photon is emitted with probability $1-\textrm{P}_e$. As detected by an ideal photon counter, this results in the photocount distribution
\begin{equation}
P_n = \{P_0, P_1, P_2, ...\} = \{1-\textrm{P}_e, \textrm{P}_e, 0, ...\},
\end{equation}
where $P_n$ is the probability to detect $n$ photons in the emitted pulse. It is on this principle that most indistinguishable single-photon sources based on solid-state quantum emitters operate\cite{schneider2015single,Fischer2016-pl}.

\begin{figure}[t]
  \vspace{0.655ex}
  \includegraphics[width=8.8cm]{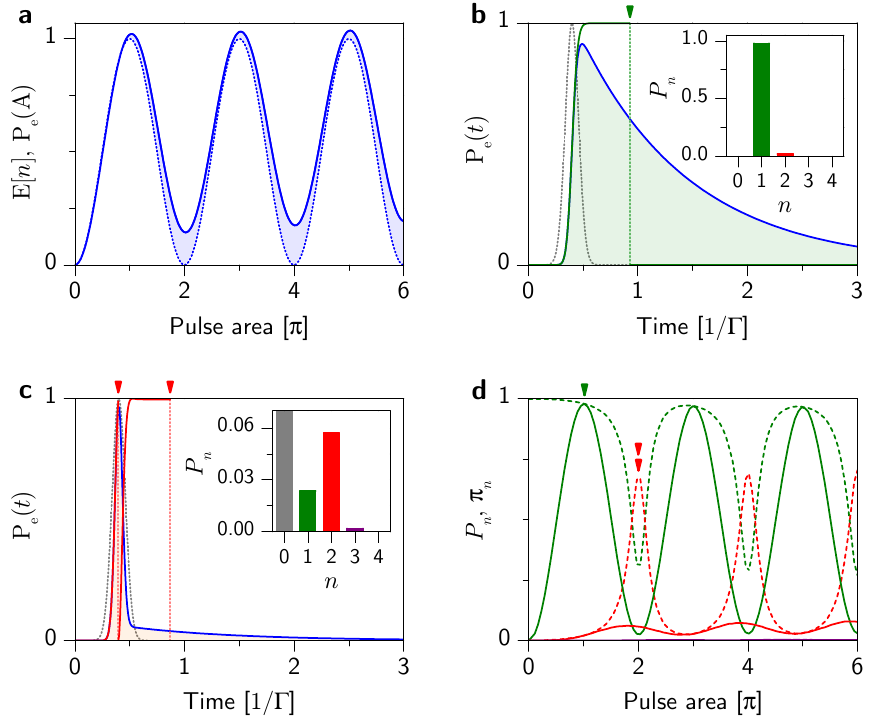}
  \caption{\textbf{Figure 1 \textbar \hspace{0.1pt} Simulations showing pulsed two-photon emission from an ideal quantum two-level system.} \textbf{a,} Rabi oscillations in excited-state population from an ideal two-level system (dashed line). Signatures of Rabi oscillations in emitted photon number under excitation with a pulse of length $\tau_\textrm{FWHM}=\tau_e/10$, where $\tau_e$ is the excited-state lifetime (solid line). \textbf{b,} System dynamics under excitation by a pulse of area~$\pi$. Dashed grey line shows driving pulse shape, blue line shows the ensemble-averaged excited-state probability, green line shows a typical quantum trajectory with one photon detection (denoted by the green triangle). Inset shows photocount histogram $P_n$ under~$\pi$-pulse excitation, with single-photon emission $P_1$ dominating. \textbf{c,} System dynamics under excitation by a pulse of area $2\pi$. Dashed grey line shows driving pulse shape, blue line shows the ensemble-averaged excited-state probability, red line shows a typical quantum trajectory with two photon emissions (denoted by the red triangles). Inset shows photocount histogram $P_n$ under $2\pi$-pulse excitation, with $P_0$ dominating but $P_2\gg P_1$. \textbf{d,} Photocount distribution $P_n$ (solid lines) and photon number purities $\pi_n$ (dashed lines) versus pulse area [under same excitation conditions as solid line in (\textbf{a})]. $n=\{1,2,3\}$ shown in colors \{green, red, and purple\}, respectively. \textit{Note: green is associated with single-photon-related indicators and red is associated with two-photon-related indicators in all figures.}
}
  \label{figure:1}
\end{figure}

A popular mechanism for approximately preparing $\ket{\psi_i}$ is the optically-driven Rabi oscillation \cite{steck2007quantum,schneider2015single}. Here, the system is initialized in its ground state and driven by a short Gaussian pulse from a coherent laser beam (of width $\tau_\textrm{FWHM}$) that is resonant with the $\ket{g}\leftrightarrow\ket{e}$ transition. Short is relative to the lifetime of the excited state $\tau_e=1/\Gamma$ in order to minimize the number of spontaneous emissions that occur during the system-pulse interaction \cite{dada2016indistinguishable,Fischer2016-pl}. As a function of the integrated pulse area, i.e. $A = {\int \mathop{\textrm{d} t} \mu\cdot E(t)/\hbar}$ where $E(t)$ is the pulse's electric field, the system undergoes coherent oscillations between its ground $\ket{g}$ and excited $\ket{e}$ states. For constant area pulses of vanishing $\tau_\textrm{FWHM}/\tau_e$, the final state of the system after interaction with the laser field is arbitrarily close to the superposition
\begin{equation}
\ket{\psi_f(A)}=\sqrt{1-\textrm{P}_e(A)}\ket{g}+\textrm{e}^{-\textrm{i}\phi}\sqrt{\textrm{P}_e(A)}\ket{e},
\end{equation}
where $\phi$ is a phase set by the laser field. Examining $\textrm{P}_e(A)$ [Fig.~\ref{figure:1}a dotted line], we see Rabi oscillations that are perfectly sinusoidal, with the laser pulse capable of inducing an arbitrary number of rotations between $\ket{g}$ and $\ket{e}$. Because $\ket{\psi_f(A)}$ looks very much like $\ket{\psi_i}$ for arbitrarily short pulses, it is commonly assumed that the photocount distribution $P_n$ always has $P_1\gg P_{n>1}$. However, we will use a quantum trajectory approach to show that, unexpectedly, $P_2>P_1$ for any $\tau_\textrm{FWHM}<\tau_e$ when $A=2n\pi$ with $n\in\{1,2,3,...\}$.

To visually illustrate the process that is capable of generating two photons, we discuss the remainder of the theory section with a convenient pulse width of $\tau_\textrm{FWHM}=\tau_e/10$. Because of the finite pulse length, we expect that in roughly $\tau_\textrm{FWHM}/\tau_e$ of the quantum trajectories a spontaneous emission occurs during the system-pulse interaction. Therefore, it is difficult to define $\ket{\psi_f}$, and the expected number of photons emitted by the system $\textrm{E}[n]$ provides a better signature of the Rabi oscillations (Fig.~\ref{figure:1}a solid line). Notably, a consequence of the spontaneous emissions is that $\textrm{E}[n]$ does not exactly follow the sinusoid of the ideal Rabi oscillations (difference highlighted with the shaded region). Because $\textrm{E}[n]>\textrm{P}_e(A)$, the system must be occasionally re-excited to emit additional photons during the system-pulse interaction.

We now examine this re-excitation process in detail, first by considering the commonly studied case of an on-demand single-photon source with $A=\pi$ (Fig.~\ref{figure:1}b). By driving a half Rabi oscillation, known colloquially as a $\pi$-pulse \cite{steck2007quantum}, the probability of single-photon generation is maximized. Because the excitation pulse is short (grey dashed line) compared to the excited-state lifetime, the emitted wavepacket has an exponential shape (blue line). To further understand the probabilistic elements of the photon emission, we study a typical quantum trajectory \cite{carmichael2009open,Johansson2013-cv} representing $\textrm{P}_e(t)$ [green line]. The system is driven by the $\pi$-pulse into its excited state, where it waits for a photon emission at some later random time (denoted by the green triangle) to return to its ground state. After computing thousands of such trajectories, $P_n$ is generated from the photodetection events and shows $P_1\approx1$ (inset) indicating that the system acts as a good single-photon source. The small amount of two-photon emission ($P_2$) occurs due to re-excitation of the quantum system during interaction with the pulse. It roughly accounts for the disparity between $\textrm{E}[n]$ and $\textrm{P}_e(A=\pi)$, and is an important but often overlooked source of error in on-demand single-photon sources. 

As our first clue that re-excitation during the system-pulse interaction can yield interesting dynamics, the difference between $\textrm{E}[n]$ and $\textrm{P}_e(A)$ is not constant as a function of $A$ and is maximized for $A=2n\pi$. Therefore, we now take a closer look at the system's dynamics for a $2\pi$-pulse (Fig.~\ref{figure:1}c) and find a photocount distribution where $P_2\gg P_1$ (inset). To understand why the two-level system counter-intuitively prefers to emit two photons over a single photon, consider a typical quantum trajectory (red line). The emission probability is proportional to $\textrm{P}_e(t)$ [blue line] which peaks halfway through the excitation pulse. Therefore, the first photon is most likely to be emitted after $\approx\pi$ of the pulse area has been absorbed (first red triangle), and a remaining $\approx\pi$ in area then re-excites the system with near unity probability to emit a second photon (second red triangle). This two-photon process is triggered during the system-pulse interaction, and although these photons are emitted within a single excited-state lifetime, they have a temporal structure known as a photon bundle\cite{munoz2014emitters}. Signatures of the bundle can be found in $\textrm{P}_e(t)$: the emission shows a peak of width $\tau_\textrm{FWHM}$ followed by a long tail of length~$\tau_e$. This shows the conditional generation of a second photon based on a first emission during the system-pulse interaction, which means that the two-photon bundling effect dominates for arbitrarily short pulses and even for long pulses as well (Supplementary Fig.~S4). Hence, although the efficiency as a pulsed two-photon source is given by $P_2\approx6\,\%$, an ideal two-level system could be operated in a much more efficient regime simply by choosing a longer pulse. We avoided this discussion in the main text because $P_3$ becomes non-negligible, which makes an intuitive interpretation of the dynamics more challenging.

To fully characterise the cross-over where $P_2>P_1$, we simulated the photocount distributions as a function of pulse area (Fig.~\ref{figure:1}d). Clear oscillations can be seen between $P_1$ and $P_2$ (solid green and red lines, respectively), and $P_2$ is out of phase from the Rabi oscillations. Notably, the oscillations in $\textrm{E}[n]$ make direct comparisons between the two probabilities difficult. To better illustrate the fraction of emission occurring by $n$-photon emission, we turn to a quantity called the photon number purity of the source \cite{munoz2014emitters}. By ignoring the vacuum component, $P_n$ is renormalized to
\begin{equation}
\pi_n=P_n/\sum_{n>0} P_n.
\end{equation}
The purities (dashed lines) very clearly oscillate between emission dominated by single-photon processes $\pi_1$ for odd-$\pi$-pulses and two-photon processes $\pi_2$ for even-$\pi$-pulses. Quite remarkably, $\pi_3$ remains negligible for \textit{all} pulse areas. Additionally, the purities reveal the limit of $\{\pi_1,\pi_2, \pi_3\}=\{0.3, 0.7, 0\}$ for arbitrarily short Gaussian pulses.

\begin{figure}[t]
  \vspace{0.655ex}
  \includegraphics[width=8.8cm]{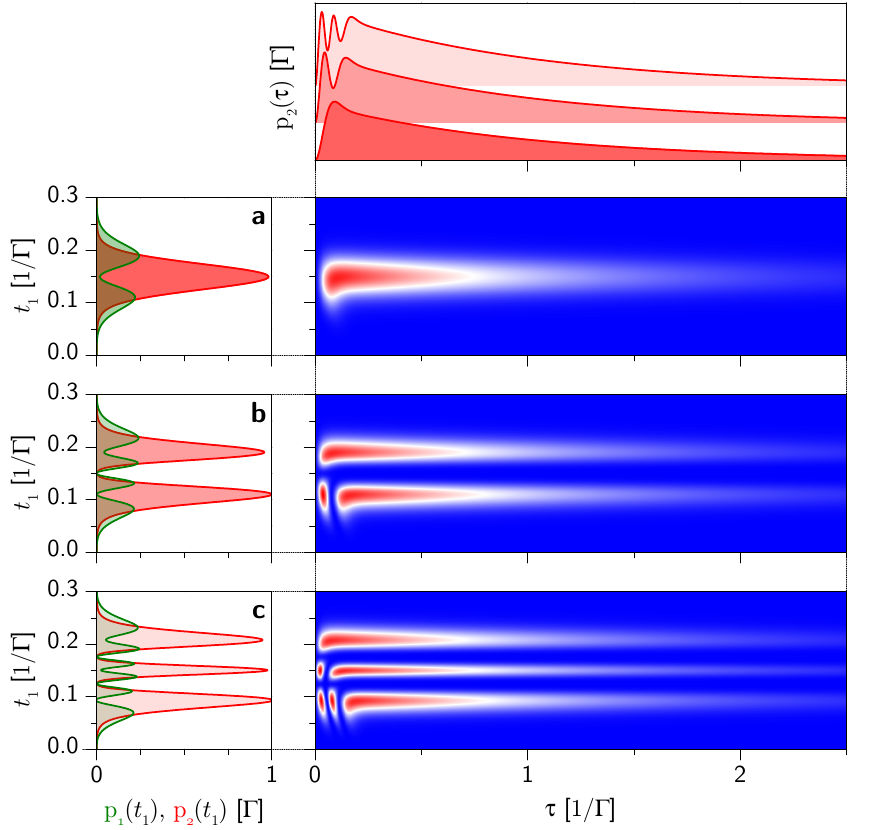}
  \caption{\textbf{Figure 2 \textbar \hspace{0.1pt} Simulations showing time-resolved single-photon and two-photon emission from an ideal quantum two-level system.} \textbf{a-c,} Probability mass functions for single-photon (green) and two-photon (red) detection showing the internal temporal structure of the photon pairs, for excitation by a $2\pi$-pulse (\textbf{a}), 4$\pi$-pulse (\textbf{b}), and 6$\pi$-pulse (\textbf{c}). Colour plots show $p_2(t_1,\tau)$, while the traces to the left show $p_1(t_1)$ and $p_2(t_1)$, and the traces to the top show $p_2(\tau)$.
}
  \label{figure:2}
\end{figure}

As suggested earlier, this two-photon emission comes as an ordered pair, where the first emission event within the pulse excitation window triggers the absorption and subsequent emission of a second photon. Unlike the first photon, the second photon has the entire excited-state lifetime to leave. This statement can be quantified by investigating the time-resolved probability mass functions for photodetection \cite{loudon2000quantum} (see Methods for a derivation from system dynamics), defined by
\begin{equation}
P_1=\int_\mathbb{R}\mathop{\textrm{d} t_1}p_1(t_1) \textrm{\quad and \quad} P_2=\iint_{\mathbb{R}^2}\mathop{\textrm{d} t_1}\mathop{\textrm{d} \tau}p_2(t_1,\tau).
\end{equation}
The mass function $p_1(t_1)$ represents the probability density for emission of a single photon at time $t_1$ with no subsequent emissions, while $p_2(t_1,\tau)$ represents the joint probability density for emission of a single photon at time $t_1$ with a subsequent emission at time $t_1+\tau$. Additionally, $p_2(t_1,\tau)$  can be integrated along $t_1$ or $\tau$ to yield $p_2(\tau)$ or $p_2(t_1)$, which give the probability density for waiting $\tau$ between the two emission events or detecting a photon pair with the first emission at time $t_1$, respectively.

We explore these temporal dynamics for excitation by a $2\pi$-, $4\pi$-, and $6\pi$-pulse in Figs.~\ref{figure:2}a, \ref{figure:2}b, and~\ref{figure:2}c, respectively. First, consider excitation by the $2\pi$-pulse: $p_2(t_1,\tau)$ captures the dynamics already discussed through having a high probability of the first emission at time $t_1$ only within the pulse window of $0.1/\Gamma$, but the second emission occurs at a delay~$\tau$ later within the spontaneous emission lifetime $\tau_e$. This effect is most clearly seen in $p_1(t_1)$, $p_2(t_1)$, and $p_2(\tau)$ [traces to the left and top of the colour plots in Fig.~\ref{figure:2}], where the density of photon pair emission being triggered at time $t_1$, i.e. $p_2(t_1)$, is maximized after $\pi$ of the pulse has been absorbed ($t_1=0.15/\Gamma$) and reaches nearly unity. The second photon of the pair then has the entire lifetime to leave, as seen in the long correlation time for $p_2(\tau)$. Meanwhile, the enhancement in photon pair production leads to a corresponding decrease in density of single-photon emissions $p_1(t_1)$ around $t_1=0.15/\Gamma$. 

Next, consider excitation by the $4\pi$-pulse: $p_2(t_1,\tau)$ shows the effects of an additional Rabi oscillation that the system undergoes during interaction with the pulse. If the first emission occurs after $\pi$ of the pulse has been absorbed, a remaining $3\pi$ can result in a second emission in two different ways: either after absorption of $\pi$ additional energy or after absorption of $3\pi$ additional energy. On the other hand, if the first emission occurs after $3\pi$ of the pulse has been absorbed, only a remaining $\pi$ can be absorbed, resulting in a monotonic region of $p_2(t_1,\tau)$ just like for the $2\pi$ case. In either scenario, the probability of two emissions is most likely (but three emissions almost never happen) because a single emission converts an even-$\pi$-pulse into an odd-$\pi$-pulse, which anti-bunches the next emission. The high fidelity of this conversion process can clearly be observed in $p_1(t_1)$ and $p_2(t_1)$. The pair production is most likely after either $\pi$ or $3\pi$ of the pulse has been absorbed (times $t_1=0.1$ and $t_1=0.3$, respectively), and it occurs with almost unity probability density. This means that if the first photon is emitted at time $t_1=0.1$ \textit{or} $t_1=0.3$, then the conditional probability to emit a second photon is near unity. As a result, $p_1(t_1)$ and $p_2(t_1)$ almost look like they were just copied a second time from the $2\pi$-pulse scenario, confirming our intuitive interpretation of the $4\pi$-pulse scenario. These ideas trivially extrapolate to the $6\pi$-pulse, where three complete Rabi oscillations occur, and the projections $p_1(t_1)$ and $p_2(t_1)$ are copied once more along $t_1$.

\begin{figure}[t]
  \vspace{0.655ex}
  \includegraphics[width=8.8cm]{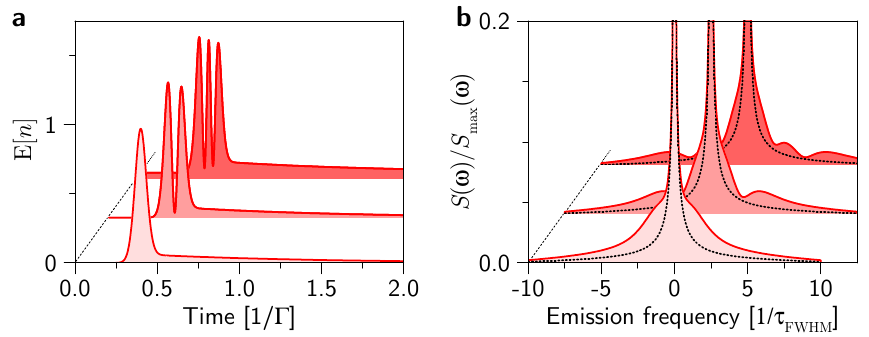}
  \caption{\textbf{Figure 3 \textbar \hspace{0.1pt} Super-natural linewidth photons from an ideal quantum two-level system.} \textbf{a,b,} Rabi oscillations (\textbf{a}) and dynamical spectra of emission (\textbf{b}) under excitation by a $2\pi$-, $4\pi$-, and $6\pi$-pulse, denoted with increasing darkness of the shaded region for higher pulse area. Dashed lines show natural emission linewidth.
}
  \label{figure:3}
\end{figure}

Looking at the oscillations in $p_2(\tau)$ for increasing pulse areas, one may notice a qualitative resemblance to the photon bunching \cite{nazir2008photon} behind a continuous-wave Mollow triplet \cite{steck2007quantum}. In fact, the underlying process where a photon emission collapses the system into its ground state, restarting a Rabi rotation, is responsible for the dynamics in both cases. However, our observed phenomenon has a very important difference: after a photodetection the expected waiting time for the second, third, and $n$-th photon emissions is identical in the continuous case, while our observed process dramatically suppresses $P_3$.

Because this method of generating two-photon bundles requires the emission of the first photon to occur within a tightly defined time interval, as set by the pulse width, we explore the emission in the context of time-frequency uncertainty (Fig.~\ref{figure:3}). First consider the $2\pi$-pulse case: we replot $\textrm{P}_e(t)$ as the light red shaded trace (a), which results in the first shaded emission spectrum \cite{wilkens1989resonance} (b). Compared to the natural linewidth of the system's transition (dashed black line), the emission is spectrally broadened by the first emission of a \textit{super}-natural linewidth photon of order $1/\tau_\textrm{FWHM}$, which occurs during the laser pulse. We note that it would be interesting to explore the physics of super-natural linewidth photons that have been \textit{incoherently} emitted as the result of a new many-body scattering process (having isolated them with a spectral notch filter to remove the second photon of a natural linewidth). As an effect of the increased number of Rabi oscillations (see for $4\pi$- and $6\pi$-pulses), the super-natural linewidth photons show oscillations in spectral power density that resemble an emerging dynamical Mollow triplet \cite{moelbjerg2012resonance}.

\begin{figure}[t]
  \vspace{0.655ex}
  \includegraphics[width=8.8cm]{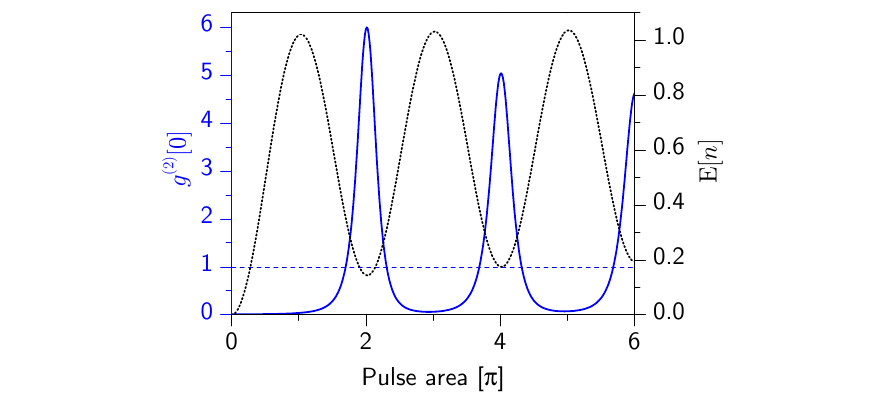}
  \caption{\textbf{Figure 4 \textbar \hspace{0.1pt} Bunched photon pair emission from an ideal quantum two-level system.} Normalised second-order factorial moment of the photocount distribution (blue), which measures the total degree of second-order coherence $g^{(2)}[0]$. Dashed blue line indicates the Poissonian counting statistics of the laser pulse. Dotted black line again indicates Rabi oscillations for reference.
}
  \label{figure:4}
\end{figure}

\begin{figure*}[th!]
  \vspace{0.655ex}
  \centering
  \includegraphics[width=17cm]{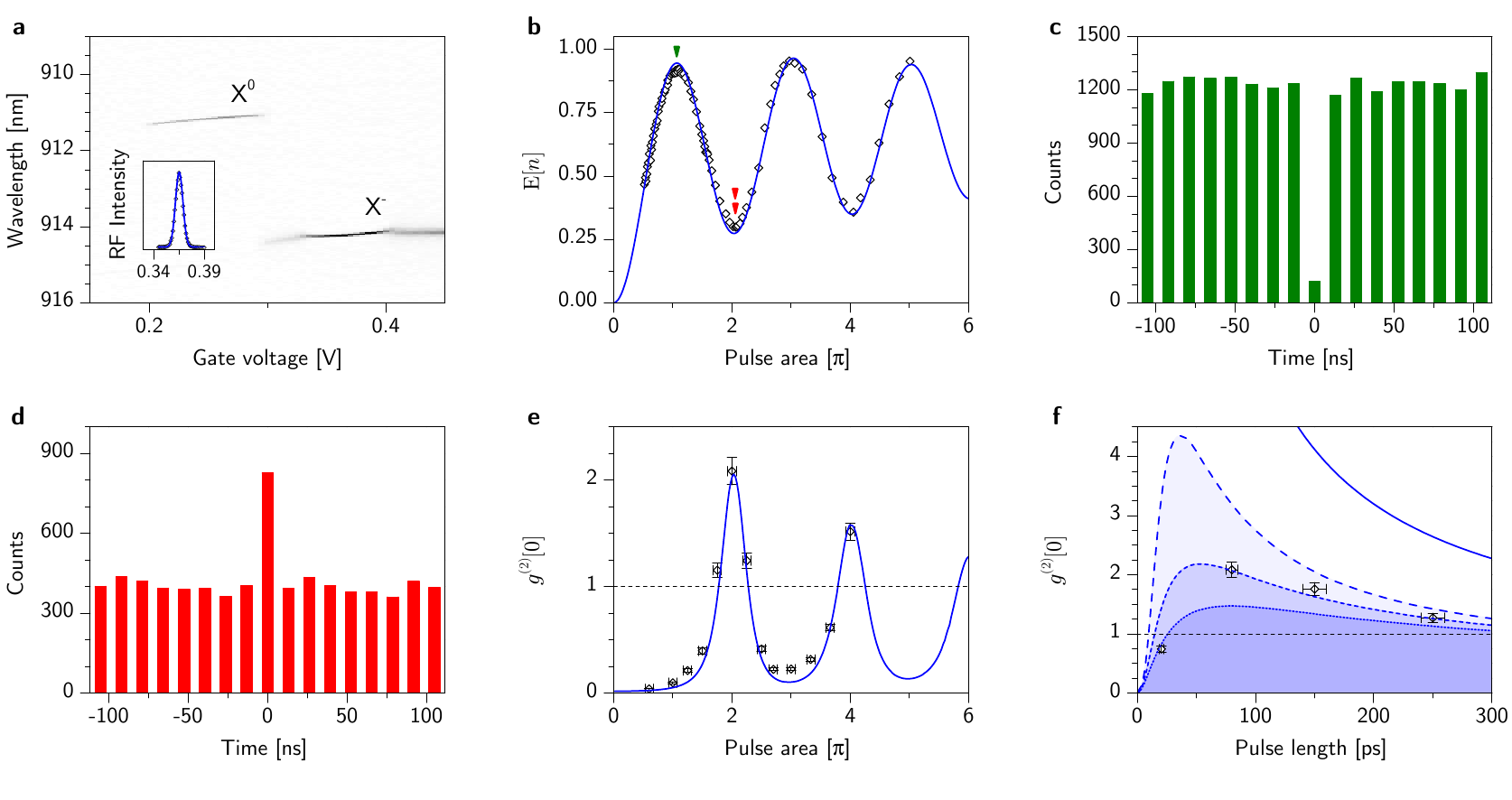}
  \caption{\textbf{Figure 5 \textbar \hspace{0.1pt} Experiments showing two-photon emission from a single artificial atom's transition.} \textbf{a,} Photoluminescence versus applied gate bias and wavelength showing the charge-stability region of the X\textsuperscript{-} transition. Inset: Voltage dependence of resonance fluorescence intensity under pulsed excitation, with optimal resonance fluorescence signal occurring at $V_g=\textsf{0.365\,V}$. \textbf{b,} Experimental resonance fluorescence signal showing Rabi oscillations, scaled to quantum simulations of emitted photon number (blue). \textbf{c,d,} Under excitation by $\pi$- and $2\pi$-pulses (green and red triangles in [\textbf{b}]), Hanbury-Brown and Twiss data respectively show anti-bunching ($g^{(2)}[0]<1$) [\textbf{c}] and bunching ($g^{(2)}[0]>1$) [\textbf{d}]. The measured values are $g^{(2)}[0]= \textsf{0.096} \pm \textsf{0.009}$ and $g^{(2)}[0] = \textsf{2.08} \pm \textsf{0.13}$, respectively. \textbf{e,} Experimental second-order coherence measurements $g^{(2)}[0]$ versus pulse area showing oscillations between anti-bunching (at odd $\pi$-pulses) and bunching (at even-$\pi$-pulses). Blue curve represents quantum simulations of the time-integrated correlations $g^{(2)}[0]$ from the experimental system. Dashed black line represents statistics of the incident laser pulse. \textbf{f,} Experimental second-order coherence measurements $g^{(2)}[0]$ versus pulse length (using $2\pi$-pulses for excitation). Optimal bunching, and hence two-photon bundling, occurs for the 80\,ps pulse. Solid blue line represents emission from an ideal two-level quantum system, long dashed blue line represents inclusion of dephasing, short dashed blue line represents addition of a 2.7\,\% chirp in bandwidth, and short dotted blue line represents addition of a further 2.7\,\% chirp in bandwidth. Again, dashed black line represents statistics of the incident laser pulse. Note, the errors in $g^{(2)}[0]$ values for (\textbf{e}) and (\textbf{f}) are the standard $\sqrt{n}$ fluctuations in the photocount distribution\cite{Fischer2016-pl}. The error in pulse area accounts for power drifts during the experiment and the errors in pulse lengths are least squares fitting errors to the pulse spectra.
}
  \label{figure:5}
\end{figure*}

Finally, we discuss the counting statistics of the emitted light in comparison to a coherent laser pulse to verify the nonclassical nature of the emission. Because the photocount distribution is fully described with just $P_0$, $P_1$, and $P_2$, its information is completely contained in the mean $\textrm{E}[n]$ and the normalised second-order factorial moment  \cite{carmichael2009open}
\begin{equation}
g^{(2)}[0] \equiv \frac{\textrm{E}[n\left(n-1\right)]}{\textrm{E}[n]^2},
\end{equation}
which physically describes the relative probability to detect a correlated photon pair over randomly finding two un-correlated photons in a Poissonian laser pulse of equivalent mean. Thus, the quantity $g^{(2)}[0]$ yields important information on how a beam of light deviates from the Poissonian counting statistics of a laser beam. For Poissonian statistics $g^{(2)}[0]=1$, but for sub-Poissonian statistics $g^{(2)}[0]<1$ (anti-bunching) and for super-Poissonian statistics $g^{(2)}[0]>1$ (bunching) \cite{Fischer2016-pl}. In particular, because the emission under a $2\pi$ excitation is a weak two-photon pulse, we expect that the photons will arrive in ``bunched'' pairs, where the first detection heralds the presence of a second photon in the pulse. This prediction is confirmed in Fig.~\ref{figure:4}, where emission for even-$\pi$-pulses strongly bunches, thus confirming the highly nonclassical nature of the emission and providing an experimentally accessible signature of the oscillations in $P_2$.


\textit{Experiment}---After having theoretically discovered that a quantum two-level system is able to preferentially emit two photons through a complex many-body scattering phenomenon, we found experimental signatures of this two-photon process using a single transition from an artificial atom. Our artificial atom of choice is an InGaAs quantum dot, due to its technological maturity and good optical quality\cite{Buckley2012-uj}. The dot is embedded within a diode structure in order to minimize charge and spin noise (Methods), resulting in a nearly transform-limited optical transition\cite{Kuhlmann2015-hh}; luminescence experiments as a function of gate voltage (Fig.~\ref{figure:5}a) reveal the charge-stability region in which we operate\cite{Warburton2000-rf} ($V_g=0.365\,\text{V}$). We used the X\textsuperscript{-} transition due to its lack of fine-structure, which results in a true two-level system (at zero magnetic field) with an excited-state lifetime of $\tau_e=602\,\text{ps}$ (Supplementary Fig.~S1). Exciting the system with laser pulses ($\tau_\textrm{FWHM}=80\,\text{ps}$), we drove Rabi oscillations between its ground and excited states (Fig.~\ref{figure:5}b). However, because the artificial atom resides in a solid-state environment, it possesses several non-idealities that slightly decrease the fidelity of the oscillations: a power-dependent dephasing rate arising from electron-phonon interaction\cite{Ramsay2010-bd} and an excited-state dephasing due to spin or charge noise\cite{Kuhlmann2015-hh}. Additionally, the quantum systems are very sensitive to minimal pulse chirps arising due to optical setup non-idealities\cite{Debnath2012-kr}. Using these three effects as fitting parameters (Methods), we obtained near perfect agreement between our quantum-optical model (blue) and the experimental Rabi oscillations.

Next, we measured the $g^{(2)}[0]$ values of the emitted wavepackets, in order to study the photon bunching effects outlined in Fig.~\ref{figure:4}. Two typical experiments are presented in Figs.~\ref{figure:5}c and \ref{figure:5}d, showing $g^{(2)}[0]\approx0$ (anti-bunching) and $g^{(2)}[0]>1$ (bunching) for $\pi$- and $2\pi$-pulses, respectively. A complete data-set is shown in Fig.~\ref{figure:5}e, with oscillations between anti-bunching at odd-$\pi$ pulses and bunching at even-$\pi$-pulses. Using the same fitting parameters as in Fig.~\ref{figure:5}b, the correlation data are almost perfectly matched with our full quantum-optical model. Hence, we have found experimental evidence that suggests the artificial atom is affected by the predicted many-body two-photon scattering process that causes $P_2$ oscillations out of phase from the Rabi oscillations.

Finally, we investigated how non-idealities affect the bunching values (Fig.~\ref{figure:5}f) by experimentally characterising the emission at four pulse lengths. From our quantum-optical model, we see that bunching is strongest for the ideal case (blue line), and decreases for every added non-ideality (long dashed blue for dephasing, short dashed blue for additional $2.7\%$ chirp in bandwidth, and short dotted blue for further $2.7\%$ chirp in bandwidth), yielding excellent agreement with the data. Discussing these effects further, an enhanced pulse chirp decreases the fidelity of photon bunching due to the function of a large chirp to adiabatically prepare the system in its excited state\cite{Ardelt2014-rq}, which decays with a single-photon emission. The minimal chirp that we observed can be removed with pulse compressors in future experiments to achieve an even better match with the ideal photon bunching curve. Additionally, at short pulse lengths the power-dependent dephasing results in anti-bunching. Due to the higher amplitude of shorter pulses (with fixed area), the dephasing rate diverges and the system acts as an incoherently pumped single-photon source. Thus, when including non-idealities we found the optimum bunching to occur at a pulse length of approximately  $80\,\text{ps}$, which indicates where the two-photon process is strongest experimentally. Although we expect the two-photon emission is dominant, the non-idealities of the solid-state system could result in non-negligible $P_3$ or higher $P_n$. This scenario is unfortunately not distinguishable through measuring $g^{(2)}[0]$ alone, but it is recently becoming possible to measure higher-order photon correlations that could help definitively identify regimes of operation where  $P_2\gg P_{3}$ \cite{Rundquist2014-kf,stevens2014third}. Finally, we expect future investigations on exploring optimal pulse shapes to enable much more efficient and higher purity two-photon emission both from ideal and experimental two-level systems.

\end{spacing}


\vspace{-1.9ex}
 
\section*{{\fontsize{11}{11}\textsf{\textbf{Methods\vspace{-1.2 ex}}}}}
{\footnotesize
\begin{spacing}{1.05}
The sample investigated is grown by molecular beam epitaxy (MBE). It consists of a layer of InGaAs quantum dots with low areal density ($<1\,\mu\text{m}^{-2}$), embedded within the intrinsic region of a Schottky photodiode formed from an n-doped layer below the quantum dots and a semitransparent titanium gold front contact. The distance between the doped layer and the quantum dots is $35\,\text{nm}$, which enables control over the charge status of the dot\cite{Warburton2000-rf}. A weak planar microcavity with an optical thickness of one wavelength ($\lambda$) is formed from a buried 18\,pair GaAs/AlAs distributive Bragg reflector (DBR) and the semitransparent top contact, which enhances the in- and out-coupling of light.

All optical measurements were performed at $4.2\,\text{K}$ in a liquid helium dipstick setup. For excitation and detection a microscope objective with a numeric aperture of $NA=0.68$ was used. Cross-polarised measurements were performed using a polarising beam splitter. To further enhance the extinction ratio, additional thin film linear polarisers were placed in the excitation/detection pathways and a single mode fibre was used to spatially filter the detection signal. Furthermore, a quarter-wave plate was placed between the beamsplitter and the microscope objective to correct for birefringence of the optics and sample itself\cite{Kuhlmann2013}. For the Rabi oscillations, a very weak laser background (due to an imperfect suppression of the excitation laser) was subtracted. This linearly increasing background was directly measured through electrically tuning the quantum dot out of resonance and typically amounted to less than 10\,\% of the signal by $5\pi$ pulse area; the quantum statistic $g^{(2)}[0]$ is dependent on the square of the signal's power, which means that the background (at a maximum) contributed to less than 1\,\% of those measurements.

The $20\,\text{ps}$ and $80\,\text{ps}$ long excitation pulses were derived from a fs-pulsed titanium sapphire laser (Coherent Mira 900) through pulse shaping. For the $20\,\text{ps}$ long pulses, a 4f pulse shaper with a focal length of $1\,\text{m}$ and an $1800\,\textrm{l/mm}$ grating was used. For the $80\,\text{ps}$ long pulses a spectrometer-like filter with a focal length of $1\,\text{m}$ and an $1800\,\textrm{l/mm}$ grating was used. Longer pulses were obtained through modulating a continuous wave laser. For the modulation, a fibre-coupled and EOM-controlled lithium niobate Mach-Zehnder (MZ) interferometer with a bandwidth of $10\,\text{GHz}$ (Photline NIR-MX-LN-10) was used. Such modulators allow control of the output intensity through a DC bias and a radio frequency input. The radio frequency pulses were generated by a $3.35\,\text{GHz}$ pulse-pattern generator (Agilent 81133A). To obtain a high extinction ratio, the temperature of the modulator was stabilised and precisely controlled ($1\,\text{mK}$) using a Peltier element, thermistor, and TEC controller. This enabled a static extinction ratio $>35\,\text{db}$.

Second-order autocorrelation measurements were performed using a Hanbury-Brown and Twiss (HBT) setup consisting of one beamsplitter and two single-photon avalanche diodes. Note: their timing resolution ($\approx 100$\,ps) is too low to measure the correlations predicted in Fig. \ref{figure:2}. The detected photons were correlated with a TimeHarp200 time counting module. The integration times were between 30 minutes and two hours.

Quantum-optical simulations were performed with the Quantum Optics Toolbox in Python (QuTiP)\cite{Johansson2013-cv}, where the standard quantum two-level system was used as a starting point. The dynamical calculations, especially those of the measured degrees of second-order coherence $g^{(2)}[0]$, were calculated using a dynamical form of the quantum regression theorem\cite{Fischer2016-pl}. The driving laser was modelled as a Gaussian pulse, where the product of the transition dipole and electric field is given by $\mu\cdot E(t)/\hbar=A/\sqrt{\tau_p^2 \pi}\textrm{e}^{-t^2/\tau_p^2}$, $\tau_p=\tau_\textrm{FWHM}/\sqrt{2 \ln{2}}$, and $A$ is the pulse area. The chirp\cite{Debnath2012-kr} was modelled by multiplying the driving field by an additional exponential $\textrm{e}^{-\textrm{i}\alpha t^2}$, where $\alpha$ is the chirp parameter. As a function of the percent change in bandwidth due to the chirp $\Delta_\textrm{BW}$, then $\alpha = \sqrt{2\Delta_\textrm{BW} + \Delta_\textrm{BW}^2}/\tau_p^2$. The phonon-induced dephasing\cite{Ramsay2010-bd} was modelled as a power-dependent collapse operator in the quantum-optical master equation, i.e. with $c_\textrm{phonon}=\sqrt{B\left(\mu\cdot E(t)/\hbar\right)^2}\sigma^\dagger \sigma$ where $\sigma$ is the atomic lowering operator and the phonon parameter was fitted to be $B=2\cdot 10^{-3}/\textrm{GHz}$. A phenomenological dephasing rate that accounted for the spin and charge noise was modelled with the collapse operator $c_\textrm{noise}=\sqrt{\gamma_d}\sigma^\dagger \sigma$, where we fitted $\gamma_d=1.3/\text{ns}$. This dephasing rate is slightly lower than the spontaneous emission rate ($\gamma_d=0.78 \Gamma$), which is consistent with state-of-the-art values for $X^-$ transitions in InGaAs quantum dots\cite{Kuhlmann2015-hh}.

Because the ideal quantum two-level system emits negligible $P_3$ for short pulses, then the probability mass function for joint photodetection at two different times is simply given by $p_2(t_1,\tau)=G^{(2)}(t_1,\tau)/2$. $G^{(2)}(t_1,\tau)$ is the standard Glauber second-order coherence function, which can be calculated for a pulsed two-level system using a time-dependent form of the quantum regression theorem \cite{Fischer2016-pl}. Next, we discuss how to obtain $p_1(t_1)$, which is slightly more nuanced. Consider a trajectory for the ideal two-level system under excitation by an even-$\pi$-pulse: $\textrm{P}_e(A)$ always returns to zero if no emission events occur during the system-pulse interaction. Therefore, the probability density of a first detection at time $t_1$ is given by $\Gamma \textrm{P}_e(t_1)$, and this density is the sum of emissions that yield only a single photon $p_1(t_1)$ and of emissions that yield two photons $p_2(t_1)$. Hence, $p_1(t_1)=\Gamma \textrm{P}_e(t_1)-p_2(t_1)$.

The data that support the plots within this paper and other findings of this study are available from the corresponding authors upon reasonable request.
\end{spacing}
}

\vspace{-0.9ex}

\bibliographystyle{naturemag}
{\footnotesize\bibliography{bibliography}}

\providecommand{\noopsort}[1]{}\providecommand{\singleletter}[1]{#1}%
\begin{thebibliography}{10}
\expandafter\ifx\csname url\endcsname\relax
  \def\url#1{\texttt{#1}}\fi
\expandafter\ifx\csname urlprefix\endcsname\relax\def\urlprefix{URL }\fi
\providecommand{\bibinfo}[2]{#2}
\providecommand{\eprint}[2][]{\url{#2}}

\bibitem{shen2007sc}
\bibinfo{author}{Shen, J.-T.} \& \bibinfo{author}{Fan, S.}
\newblock \bibinfo{title}{Strongly correlated two-photon transport in a
  one-dimensional waveguide coupled to a two-level system}.
\newblock \emph{\bibinfo{journal}{Phys. Rev. Lett.}}
  \textbf{\bibinfo{volume}{98}}, \bibinfo{pages}{153003}
  (\bibinfo{year}{2007}).

\bibitem{zheng2010waveguide}
\bibinfo{author}{Zheng, H.}, \bibinfo{author}{Gauthier, D.~J.} \&
  \bibinfo{author}{Baranger, H.~U.}
\newblock \bibinfo{title}{Waveguide {QED}: Many-body bound-state effects in
  coherent and fock-state scattering from a two-level system}.
\newblock \emph{\bibinfo{journal}{Phys. Rev. A}} \textbf{\bibinfo{volume}{82}},
  \bibinfo{pages}{063816} (\bibinfo{year}{2010}).

\bibitem{pletyukhov2015quantum}
\bibinfo{author}{Pletyukhov, M.} \& \bibinfo{author}{Gritsev, V.}
\newblock \bibinfo{title}{Quantum theory of light scattering in a
  one-dimensional channel: Interaction effect on photon statistics and
  entanglement entropy}.
\newblock \emph{\bibinfo{journal}{Phys. Rev. A}} \textbf{\bibinfo{volume}{91}},
  \bibinfo{pages}{063841} (\bibinfo{year}{2015}).

\bibitem{schneider2015single}
\bibinfo{author}{Schneider, C.} \emph{et~al.}
\newblock \bibinfo{title}{Single semiconductor quantum dots in microcavities:
  bright sources of indistinguishable photons}.
\newblock In \emph{\bibinfo{booktitle}{Engineering the Atom-Photon
  Interaction}}, \bibinfo{pages}{343--361} (\bibinfo{publisher}{Springer},
  \bibinfo{year}{2015}).

\bibitem{Fischer2016-pl}
\bibinfo{author}{Fischer, K.~A.}, \bibinfo{author}{M{\"u}ller, K.},
  \bibinfo{author}{Lagoudakis, K.~G.} \& \bibinfo{author}{Vu{\v{c}}kovi{\'c},
  J.}
\newblock \bibinfo{title}{Dynamical modeling of pulsed two-photon
  interference}.
\newblock \emph{\bibinfo{journal}{New J. Phys.}} \textbf{\bibinfo{volume}{18}},
  \bibinfo{pages}{113053} (\bibinfo{year}{2016}).

\bibitem{pletyukhov2013full}
\bibinfo{author}{Pletyukhov, M.}, \bibinfo{author}{Ringel, M.} \&
  \bibinfo{author}{Gritsev, V.}
\newblock \bibinfo{title}{Full counting statistics of photons interacting with
  emitter}  (\bibinfo{year}{2013}).
\newblock \eprint{arXiv:1302.2156}.

\bibitem{lindkvist2014scattering}
\bibinfo{author}{Lindkvist, J.} \& \bibinfo{author}{Johansson, G.}
\newblock \bibinfo{title}{Scattering of coherent pulses on a two-level
  system---single-photon generation}.
\newblock \emph{\bibinfo{journal}{New J. Phys.}} \textbf{\bibinfo{volume}{16}},
  \bibinfo{pages}{055018} (\bibinfo{year}{2014}).

\bibitem{carmichael2009open}
\bibinfo{author}{Carmichael, H.}
\newblock \emph{\bibinfo{title}{An open systems approach to quantum optics:
  lectures presented at the Universit{\'e} Libre de Bruxelles, October 28 to
  November 4, 1991}}, vol.~\bibinfo{volume}{18} (\bibinfo{publisher}{Springer
  Science \& Business Media}, \bibinfo{year}{2009}).

\bibitem{dada2016indistinguishable}
\bibinfo{author}{Dada, A.~C.} \emph{et~al.}
\newblock \bibinfo{title}{Indistinguishable single photons with flexible
  electronic triggering}.
\newblock \emph{\bibinfo{journal}{Optica}} \textbf{\bibinfo{volume}{3}},
  \bibinfo{pages}{493--498} (\bibinfo{year}{2016}).

\bibitem{Cohen-Tannoudji1992-uo}
\bibinfo{author}{Cohen-Tannoudji, C.}, \bibinfo{author}{Dupont-Roc, J.},
  \bibinfo{author}{Grynberg, G.} \& \bibinfo{author}{Thickstun, P.}
\newblock \emph{\bibinfo{title}{Atom-photon interactions: basic processes and
  applications}} (\bibinfo{publisher}{Wiley Online Library},
  \bibinfo{year}{1992}).

\bibitem{steck2007quantum}
\bibinfo{author}{Steck, D.~A.}
\newblock \emph{\bibinfo{title}{Quantum and atom optics}}
  (\bibinfo{year}{2007}).

\bibitem{Johansson2013-cv}
\bibinfo{author}{Johansson, J.~R.}, \bibinfo{author}{Nation, P.~D.} \&
  \bibinfo{author}{Nori, F.}
\newblock \bibinfo{title}{{QuTiP} 2: A python framework for the dynamics of
  open quantum systems}.
\newblock \emph{\bibinfo{journal}{Comput. Phys. Commun.}}
  \textbf{\bibinfo{volume}{184}}, \bibinfo{pages}{1234--1240}
  (\bibinfo{year}{2013}).

\bibitem{munoz2014emitters}
\bibinfo{author}{Mu{\~n}oz, C.~S.} \emph{et~al.}
\newblock \bibinfo{title}{Emitters of n-photon bundles}.
\newblock \emph{\bibinfo{journal}{Nat. Photon.}} \textbf{\bibinfo{volume}{8}},
  \bibinfo{pages}{550--555} (\bibinfo{year}{2014}).

\bibitem{loudon2000quantum}
\bibinfo{author}{Loudon, R.}
\newblock \emph{\bibinfo{title}{The quantum theory of light}}
  (\bibinfo{publisher}{OUP Oxford}, \bibinfo{year}{2000}).

\bibitem{nazir2008photon}
\bibinfo{author}{Nazir, A.}
\newblock \bibinfo{title}{Photon statistics from a resonantly driven quantum
  dot}.
\newblock \emph{\bibinfo{journal}{Phy. Rev. B}} \textbf{\bibinfo{volume}{78}},
  \bibinfo{pages}{153309} (\bibinfo{year}{2008}).

\bibitem{wilkens1989resonance}
\bibinfo{author}{Wilkens, M.} \& \bibinfo{author}{Rza̧ewski, K.}
\newblock \bibinfo{title}{Resonance fluorescence of an arbitrarily driven
  two-level atom}.
\newblock \emph{\bibinfo{journal}{Phys. Rev. A}} \textbf{\bibinfo{volume}{40}},
  \bibinfo{pages}{3164} (\bibinfo{year}{1989}).

\bibitem{moelbjerg2012resonance}
\bibinfo{author}{Moelbjerg, A.}, \bibinfo{author}{Kaer, P.},
  \bibinfo{author}{Lorke, M.} \& \bibinfo{author}{M{\o}rk, J.}
\newblock \bibinfo{title}{Resonance fluorescence from semiconductor quantum
  dots: beyond the mollow triplet}.
\newblock \emph{\bibinfo{journal}{Phys. Rev. Lett.}}
  \textbf{\bibinfo{volume}{108}}, \bibinfo{pages}{017401}
  (\bibinfo{year}{2012}).

\bibitem{Buckley2012-uj}
\bibinfo{author}{Buckley, S.}, \bibinfo{author}{Rivoire, K.} \&
  \bibinfo{author}{Vu\v{c}kovi\'{c}, J.}
\newblock \bibinfo{title}{Engineered quantum dot single-photon sources}.
\newblock \emph{\bibinfo{journal}{Rep. Prog. Phys.}}
  \textbf{\bibinfo{volume}{75}}, \bibinfo{pages}{126503}
  (\bibinfo{year}{2012}).

\bibitem{Kuhlmann2015-hh}
\bibinfo{author}{Kuhlmann, A.~V.} \emph{et~al.}
\newblock \bibinfo{title}{Transform-limited single photons from a single
  quantum dot}.
\newblock \emph{\bibinfo{journal}{Nat. Commun.}} \textbf{\bibinfo{volume}{6}},
  \bibinfo{pages}{8204} (\bibinfo{year}{2015}).

\bibitem{Warburton2000-rf}
\bibinfo{author}{{Warburton}} \emph{et~al.}
\newblock \bibinfo{title}{Optical emission from a charge-tunable quantum ring}.
\newblock \emph{\bibinfo{journal}{Nature}} \textbf{\bibinfo{volume}{405}},
  \bibinfo{pages}{926--929} (\bibinfo{year}{2000}).

\bibitem{Ramsay2010-bd}
\bibinfo{author}{Ramsay, A.~J.} \emph{et~al.}
\newblock \bibinfo{title}{{Phonon-Induced} {Rabi-Frequency} renormalization of
  optically driven single {InGaAs} / {GaAs} quantum dots}.
\newblock \emph{\bibinfo{journal}{Phys. Rev. Lett.}}
  \textbf{\bibinfo{volume}{105}}, \bibinfo{pages}{177402}
  (\bibinfo{year}{2010}).

\bibitem{Debnath2012-kr}
\bibinfo{author}{Debnath, A.}, \bibinfo{author}{Meier, C.},
  \bibinfo{author}{Chatel, B.} \& \bibinfo{author}{Amand, T.}
\newblock \bibinfo{title}{Chirped laser excitation of quantum dot excitons
  coupled to a phonon bath}.
\newblock \emph{\bibinfo{journal}{Phys. Rev. B}} \textbf{\bibinfo{volume}{86}},
  \bibinfo{pages}{161304} (\bibinfo{year}{2012}).

\bibitem{Ardelt2014-rq}
\bibinfo{author}{Ardelt, P.-L.} \emph{et~al.}
\newblock \bibinfo{title}{Dissipative preparation of the exciton and biexciton
  in self-assembled quantum dots on picosecond time scales}.
\newblock \emph{\bibinfo{journal}{Phys. Rev. B}} \textbf{\bibinfo{volume}{90}},
  \bibinfo{pages}{241404} (\bibinfo{year}{2014}).

\bibitem{Rundquist2014-kf}
\bibinfo{author}{Rundquist, A.} \emph{et~al.}
\newblock \bibinfo{title}{Nonclassical higher-order photon correlations with a
  quantum dot strongly coupled to a photonic-crystal nanocavity}.
\newblock \emph{\bibinfo{journal}{Phys. Rev. A}} \textbf{\bibinfo{volume}{90}},
  \bibinfo{pages}{023846} (\bibinfo{year}{2014}).

\bibitem{stevens2014third}
\bibinfo{author}{Stevens, M.~J.}, \bibinfo{author}{Glancy, S.},
  \bibinfo{author}{Nam, S.~W.} \& \bibinfo{author}{Mirin, R.~P.}
\newblock \bibinfo{title}{Third-order antibunching from an imperfect
  single-photon source}.
\newblock \emph{\bibinfo{journal}{Opt. Express}} \textbf{\bibinfo{volume}{22}},
  \bibinfo{pages}{3244--3260} (\bibinfo{year}{2014}).

\bibitem{Kuhlmann2013}
\bibinfo{author}{Kuhlmann, A.~V.} \emph{et~al.}
\newblock \bibinfo{title}{A dark-field microscope for background-free detection
  of resonance fluorescence from single semiconductor quantum dots operating in
  a set-and-forget mode}.
\newblock \emph{\bibinfo{journal}{Rev. Sci. Instrum.}}
  \textbf{\bibinfo{volume}{84}}, \bibinfo{pages}{073905}
  (\bibinfo{year}{2013}).

\end{thebibliography}

\vspace{-3.3ex}

\section*{{\fontsize{11}{11}\textsf{\textbf{Acknowledgements\vspace{-1.2ex}}}}}
{\footnotesize \begin{spacing}{1.05}The authors thank Alex Rasmussen for productive and helpful discussions in framing the context of this work.

We gratefully acknowledge financial support from the National Science Foundation (Division of Materials Research - Grant. No. 1503759), the DFG via the Nanosystems Initiative Munich, the BMBF via Q.com (Project No. 16KIS0110) and BaCaTeC. K.A.F. acknowledges support from the Lu Stanford Graduate Fellowship and the National Defense Science and Engineering Graduate Fellowship. J.W. acknowledges support from the PhD programme ExQM of the Elite Network of Bavaria. C.D. acknowledges support from the Andreas Bechtolsheim Stanford Graduate Fellowship. J.V. gratefully acknowledges support from the TUM Institute of Advanced Study.\end{spacing}
}

\newpage

\section*{{\fontsize{11}{11}\textsf{\textbf{Author contributions\vspace{-1.2ex}}}}}
{\footnotesize \begin{spacing}{1.05}K.A.F. performed the theoretical work and modelling. L.H., T.S., J.W., and K.M. performed the experiments. C.D. performed trial experiments. J.V. and J.J.F. provided expertise. K.M. organized the collaboration and supervised the experiments. All authors participated in the discussion and understanding of the results.\end{spacing}
}

\vspace{-1.5ex}

\section*{{\fontsize{11}{11}\textsf{\textbf{Additional information\vspace{-1.2ex}}}}}
{\footnotesize \begin{spacing}{1.05}Correspondence and requests for materials should be addressed to K.A.F. and K.M.\end{spacing}
}

\vspace{-1.5ex}

\section*{{\fontsize{11}{11}\textsf{\textbf{Competing financial interests\vspace{-1.2ex}}}}}
{\footnotesize \begin{spacing}{1.05}The authors declare no competing financial interests.\end{spacing}
}

\end{document}